\documentclass{aa}  
\usepackage{graphicx}

\def\la{\raise.5ex\hbox{$<$}\kern-.8em\lower 1mm\hbox{$\sim$}}
\def\ma{\raise.5ex\hbox{$>$}\kern-.8em\lower 1mm\hbox{$\sim$}}

\def\kms{$\rm km\, s^{-1}$}
\def\cm3{$\rm cm^{-3}$}
\def\Ts{$\rm T_{*}$}
\def\Vs{$\rm V_{s}$}
\def\n0{$\rm n_{0}$}
\def\B0{$\rm B_{0}$}

\def\Te{$\rm T_{e}$}

\def\erg{$\rm erg\, cm^{-2}\, s^{-1}$}
\def\mum{$\mu$m}
\def\agr{a$_{gr}$}

\def\Hb{H$\beta$}
\usepackage{txfonts}
\begin{document}
   \title{Gas and dust spectra of the D' type symbiotic star
HD330036}

   \subtitle{}

   \author{R. Angeloni
          \inst{1,2},
          M. Contini\inst{2,1},
          S. Ciroi\inst{1}
          \and
          P. Rafanelli\inst{1}
          }

   \offprints{R. Angeloni}

   \institute{Dipartimento di Astronomia, Universit\`a di Padova, Vicolo
dell'Osservatorio 2, I-35122 Padova, Italy\\
              \email{rodolfo.angeloni@unipd.it, stefano.ciroi@unipd.it, piero.rafanelli@unipd.it}
         \and
             School of Physics and Astronomy, Tel Aviv University, Tel Aviv
69978, Israel\\
             \email{contini@post.tau.ac.il}}

   \date{Received - ; accepted -}

 
  \abstract
   {}
   {We present a comprehensive and self-consistent modelling of the D' type symbiotic star (SS) HD330036 from radio to UV.}
   {Within a colliding-wind scenario, we analyse the continuum, line and dust spectra by means of SUMA, a code that simulates the physical conditions of an emitting gaseous cloud under the coupled effect of ionization from an external radiation source and shocks.}
   {We find that the UV lines are emitted from high density gas between the
stars downstream of the reverse shock, while the optical lines are emitted downstream of
the shock propagating outwards the system. As regards with the continuum SED, three shells are identified in the IR, at 850K, 320 K and 200 K  with radii  r = 2.8 10$^{13}$ cm, 4 10$^{14}$ cm, and  10$^{15}$ cm, respectively, adopting a distance to Earth d=2.3 kpc: interestingly, all these shells appear to be circumbinary. The analysis of the unexploited ISO-SWS spectrum reveals that both PAHs and crystalline silicates coexist in HD330036, with PAHs associated to the internal shell at 850 K, and crystalline silicates stored into the cool shells at 320 K and 200 K. Strong evidence that crystalline silicates are shaped in a disk-like structure is derived on the basis of the relative band strengths. Finally, we suggest that shocks can be a reliable mechanism in activating the annealing and the consequent crystallization processes.}
{We show that a consistent interpretation of gas and dust spectra emitted by SS can be obtained by models which accounts for the coupled effect of the photoionizing flux and of shocks. The VLTI/MIDI proposal recently accepted by ESO aims to verify and better constrain some of our results by means of IR interferometric observations.}

   \authorrunning{Angeloni et al.}
   \titlerunning{Gas and dust spectra of HD330036}
   \maketitle
%

\section{Introduction}
Symbiotic systems (SS) are interacting binaries composed by a hot star,
generally a white dwarf (WD), a cool giant, and emitting nebul{\ae} created by both the photoionizing flux from the WD and by collision of the winds. 
On the basis of the near-infrared (NIR) colours, a classification
in two types, S and D, was proposed originally by Webster \& Allen (1975) according to
whether the cool star (S type) or dust (D type) dominate the 1-4 $\mu$m spectral range. Some
years later Allen (1982) introduced a third class, designated as D' type, characterised by
very red colours in the far infrared (FIR) and by a cool star of spectral type F or G, in
contrast to ordinary symbiotics where the cool giant is a M-type star (S type) or a Mira variable (D type). The D' types are quite rare objects, and nowadays only eight out of about 200 SS are known to belong to this class (Belczy{\'n}ski et al. 2000).

In the last years several studies allowed to highlight the distinctive features of D' types: for instance, both Smith et al. (2001) and Munari et al. (2001) showed that these stars display enhancements of the s-process elements that are synthesised via slow neutron captures during stellar evolution along the asymptotic giant branch (AGB). 
Moreover, Pereira et al. (2005, hereafter P05) and Zamanov et al. (2006,
hereafter Z06), measuring the rotational velocity of the cool star, pointed out that in the D'
type SS the cool component rotates faster than the isolated giants, at a substantial fraction of
the critical velocity (e.g. {\it v\,sin\,i $\sim$0.6 v$_{crit}$} in the case of HD330036).
This high rotational velocity can result in large mass loss rates, likely enhanced in the equatorial regions, and inevitably will affect the dusty environment (Soker 2002) by leading to a disk-like circumbinary structure in which the high gas density enhances dust formation and growth. It would be the dust temperature stratification in such a disk, already noticed in some D' SS
(e.g. V417 Cen, Van Winckel et al. 1994), to be at the origin of the observed IR excess. \\
Furthermore, under the hypothesis that the D' type orbits are synchronised, Z06 argue that the orbital periods would be relatively short (4-60 days) and the interbinary distance about 2 - 5 times the cool star radius. 

One of the most intriguing aspects of the symbiotic phenomenon in these cases pertains to the dusty environment. As a matter of fact these systems show a broad IR excess which,
since the first IR surveys, has been attributed to emission from circumstellar
dust. While in D type objects the dust excesses have colour temperatures near 1000 K
(Feast et al. 1983), in the D' SS  their presence is  revealed only  beyond $\sim$ 3 \mum, suggesting a lower temperature, which Allen (1984) stated to be not higher than 500 K.\\
Whichever symbiotic type (D or D'), thanks to development of IR astronomy it was  soon realized that it was difficult to explain the observed IR spectrum  by a single temperature component, and
theoretical models too, confirmed that several ''dust'' temperatures  should be combined  in order to reproduce the NIR-MIR data (e.g. Anandarao et al. 1988, Schild et al. 2001, Angeloni et al. 2007, in preparation). 

As with regards to the emission line spectra, D' types SS closely resemble planetary nebul{\ae} (PN), leading to a long controversy about the exact evolutionary status of these stars. It is noteworthy that D' types were even classified by some authors as young, compact PN with a binary core (Feibelman 1983, 1988; Lutz 1984, hereafter L84; Bhatt 1989, van Winckel et al. 1994, Corradi et al. 1999). However, Schmeja \& Kimeswenger (2001) pointed out that the NIR colours provide a reliable tool to distinguish symbiotic from genuine PN. 
Finally, based on the ongoing interaction between the cool giant and the nebula, Schmid \& Nussbaumer (1993, hereafter SN93) favour a classification of D' type systems as symbiotic stars.\\

Among D' type SS one of the most intriguing object is HD330036 (CN 1-1). Unfortunately, many physical parameters of this enigmatic system remain inconclusive. \\
The estimate of the hot star temperature, for instance, ranges from 60000 K (SN93), $10^5$ K  (L84), up to $2.5\,10^5$ K (Bhatt \& Mallik 1986, hereafter BM86).
The interpretation of polarisation measurements is also uncertain: as a matter of fact Schulte-Labdeck \& Magalhaes (1987) considered the polarisation observed ($\sim$ 3\%) in HD330036 of purely interstellar origin; whereas Bhatt (1989) argued that, at least to some extent, the polarisation can be intrinsic to the system and due to scattering by dust grains in an asymmetric nebula, calling for a bipolar morphology.\\
More debatable is the distance to Earth, an important parameter in the context of this paper. L84 estimated a distance of $\sim$450 pc based upon the colour excess versus distance for stars within 20' of HD330036, but based on the cool star luminosity there are several arguments that led P05 to assume a distance  of 2.3 kpc; the difference between the estimates of L84 and P05 being in someway caused by different values of the reddening. Summarising, the probable limits for the distance to HD330036 lie within $\sim$0.6 to 2.3 kpc, with upper values being more likely.\\
On the other hand, the cool star is rather well known despite its evolutionary status is still controversial: P05 obtained $L=650L_{\odot}$, $T_{eff}$=6200$\pm$150K, log g=2.4$\pm$0.7 where L
is the luminosity, $T_{eff}$ the effective temperature, g the gravity; this implies $R_g$=22$R_{\odot}$, $M_g$=4.46$M_{\odot}$ (using $R_g$ and log g), and $P_{rot} \leq$10.4$\pm$2.4 d, where $R_g$, $M_g$ and $P_{rot}$ stand for radius, mass and rotational period, respectively.

These parameters would be theoretically sufficient for an estimate of the mass loss rate (not found in the current literature); but the problem is to understand if the formul\ae{} for dust-enshrouded red giants and supergiants and oxygen-rich AGB stars remain valid when extended to a G-F giant. As a matter of fact, when we attempt to calculate \.{M} by using several proposed relations (Wachter et al. 2002, van Loon et al. 2005) and assuming the Pereira's stellar parameters, we find discordant results. Furthermore these values are too low (\.M$<10^{-9} M_{\odot}/yr$) in order to sustain any interaction of the binary stars that must be at the origin of the observed symbiotic activity (Kenyon 1988). Unfortunately, modelling of motions in the atmospheres of yellow giants and supergiants only managed to emphasise that the subject is still not well understood, resulting in the lack of reliable empirical mass loss rates or terminal wind velocities for normal G-F giants and supergiants (Achmad et al. 1997). Hence in this paper we decided to attempt a completely alternative approach (Kemper et al. 2001) deriving the mass loss rate by means of the crystalline dust feature recognizable in the infrared spectrum (see $\S$5.3.4).

Concerning the dusty environment of HD330036, Allen (1984) realized its uniqueness among the known symbiotic stars in exhibiting infrared emission bands at 3.3 and 11.3 $\mu$m, suggesting  a C-rich nature of this object. A few years later BM86, on the basis of IRAS observations, noticed  that there were two distinct components of infrared emitting dust in HD330036: one at a temperature  of $\sim$ 215K and the other much hotter  at $\sim$ 850K; interestingly, in order to obtain a likely dust to gas ratio, these authors postulated the existence of large dust grains (\agr $>$ 1 \mum).\\

In the present paper we aim to model HD330036 in the light of the nowadays widely accepted interpretation of SS as colliding-wind binary systems by combining UV and optical observations (reported in L84 and SN93) together with the IR ISO-SWS (Short Wavelength Spectrograph) spectrum, analysed here for the first time.
The observed line ratios allow us to constrain the physical conditions in the  emitting nebulae, while the ISO data, combined with other IR photometry points from IRAS and 2MASS,  reveal the properties of dust by constraining temperature, size and chemical composition of the HD330036 dusty
environment. \\
We start by analysing HD330036 UV and optical line spectra in $\S$3. Subsequently, cross-checking the  continuum and  line ratio results, we select the models which best
explain the gas and dust emission. We then derive the grain conditions and location by comparing the dust reprocessed radiation flux with the IR data ($\S$4). In $\S$5  we review the main characteristics of dust features by carefully analysing the ISO-SWS spectrum. Discussion and concluding remarks follow in $\S$6.


\section{The models}
\subsection{The colliding-wind scenario}
In the past years, theoretical models (Girard \& Willson 1987, Kenny \& Taylor 2005) as well as observations (Nussbaumer et al. 1995) have categorically shown that in SS both the hot and cool stars lose mass through stellar winds which collide within and outside the system, hence creating a complex network of wakes and shock fronts which result in a complicated structure of gas and dust nebul\ae{} (Nussbaumer 2000).\\
In this paper, as previously done for other SS (e.g. Angeloni et al. 2007a), we refer to two main shocks: the shock between the stars facing the WD, which is a head-on shock (hereafter the \textit{reverse} shock), and the head-on-back shock, which propagates outwards the system (hereafter the \textit{expanding} shock). Both the nebulae downstream of the shock fronts are ionized and heated by the radiation flux from the hot star and by shocks. The photoionizing radiation flux reaches the very shock front of the reverse shock, while it reaches the edge opposite to the shock front downstream of the expanding shock. This scenario is even complicated in D' type systems  by the giant fast rotation which leads to extended disk-like structures, both predicted (Z06) and
in some objects even optically resolved (van Winckel et al. 1994). \\ 
The optical spectrum contains several forbidden lines whose ratios constrain the models. The characteristic electron densities indicate that the region where these lines arise from is essentially different from the region which emits the UV lines. Thus we suggest that the optical spectrum results from the collision of the cool component wind with the ISM, in the external region  of the disk or even outside, most probably  throughout jets.
This hypothesis will be tested by  modelling the  spectra.
UV lines  corresponding  to high densities ($> 10^6$ $\rm cm^{-3}$)
are generally emitted from the nebula downstream of the reverse
shock between the stars (e.g. Contini \& Formiggini 2003, Angeloni et al. 2007a).
\subsection{The SUMA code}
The results presented in this work are performed by SUMA (Viegas \& Contini 1994; Contini 1997), a code that simulates the physical conditions of an emitting gaseous cloud under the coupled effect of ionization from an external radiation source and shocks, and in which both line and continuum emission from gas are calculated consistently with dust reprocessed radiation (grain heating and sputtering processes are also included). The derived models have been successfully applied to several SS, e.g. AG Peg (Contini 1997, 2003), HM Sge (Formiggini, Contini \& Leibowitz 1995), RR Tel (Contini \& Formiggini 1999), He2-104 (Contini \& Formiggini 2001), R Aqr (Contini \& Formiggini 2003), H1-36 (Angeloni et al. 2007b), as well as to nova stars (V1974, Contini et al. 1997 - T Pyx, Contini \& Prialnik 1997) and supernova remnants (e.g. Kepler's SNR, Contini 2004).\\

The calculations start with gas and dust entering the shock front in a steady state regime: the gas is adiabatically compressed and thermalized throughout the shock front. In the downstream region the compression is derived by solving the Rankine-Hugoniot equations (Cox 1972): the downstream region is automatically divided in plane parallel slabs in order to calculate as smoothly as possible the physical conditions throughout the nebula. Radiation transfer and optical depths of both continuum and lines are calculated for a steady state: in particular, radiation transfer of the diffused radiation is taken into account following Williams (1967). The fractional abundance of the ions in different ionization stages is calculated in each slab by solving the ionization equilibrium equations for the elements H, He, C, N, O, Ne, Mg, Si, S, Cl, Ar, and Fe.
The electron temperature in each slab is obtained from the energy equation when collisional processes prevail and by thermal balancing when radiation processes dominate.\\
Compression downstream strongly affects the gas cooling rate by free-free, free-bound, and line emission: consequently, the emitting gas will have different physical conditions depending on the shock velocity and on the pre-shock density.

Dust is included in the calculations, too. Dust and gas are coupled throughout the shock-front and downstream by the magnetic field. In each slab the sputtering of the grains is calculated, leading to grain sizes which depend on the shock velocity and on the gas density. The temperature of the grains, which depends on the grain radius, is then calculated by radiation heating from the external (primary) source and by diffuse (secondary) radiation, as well as by gas collisional heating. The dust reprocessed radiation flux is calculated by the Plank-averaged absorption coefficient of dust in each slab, and integrated throughout the nebula downstream. 

The input parameters which characterise the shock are the shock velocity, \Vs, the
preshock density of the gas, \n0, and the preshock magnetic field, \B0. The radiation
flux is determined by the temperature of the star, interpreted as a colour temperature, \Ts,
and by the ionization parameter, $U$. The dust-to-gas ratio, $d/g$ is also accounted for, as well as the relative abundances of the elements to H. 

A detailed description of the updated code is to be presented in Contini \& Viegas (2007 in preparation).
\section{The line spectra}
Low resolution ($\sim$ 7 \AA) IUE spectra were taken in 1979 September, 1980 June, and 1981 April (L84), while optical spectra were obtained on the 4m and 1m telescopes at Cerro Tololo Inter-American Observatory (CTIO) during 1977, 1978, and 1979 (L84). The IUE observational data  by SN93 were taken in 1984, July 26. \\
The observed UV and optical lines are shown in Table \ref{tab:tbl1}.
\begin{table*}
\centering
\caption{The UV and optical emission lines\label{tab:tbl1}.}
\small{
\begin{tabular}{cccccccccccccc}
\hline \hline line & obs$^1$ &  m1 & obs$^2$  & m2 &m3  &line & obs$^1$ & m4 & m5 & m6\\
\hline
 NV 1239 & 1.46$\pm$0.32 & 1.68 & -    & 0.04 &0.09& [OII] 3727+ & 10$\pm$5 & 8.
& 4.&6.3\\
 NIV 1468& 6.02$\pm$0.81 & 7.66& 4.3& 5.8&7.3  & [NeIII] 3869+ & 200$\pm$50 &
197. & 140.&155.\\
 CIV 1548& 48.9$\pm$4.57 & 44.8 & 46.  & 40.0&44.7& [CIII] 4068   & $< 5
\pm$4&0.07& 0.07&3$^5$     \\
 HeII 1640& 1.65$\pm$0.25 & 2.28& 1.1: &0.9 &0.3& HI 4340 & 40$\pm$11& 46.& 45.&
45. \\
 OIII] 1662& 5.61$\pm$0.76 & 4.8& 5.3&4.4  &4.25& [OIII] 4363 & 70$\pm$18 & 50.&
71.& 84.\\
 NIII] 1750& 6.12$\pm$0.71 & 5.0 & 5.7&4.8&5.03  & HeI 4471 & $<5 \pm$4 & 4.4&
5.7& 6.\\
 CIII] 1911& 19.11$\pm$1.51& 19.6& 21. & 19.2&19.2 & HeII 4686 & 5$\pm$4 & 18.&
9.&6.3 \\
 \Hb 4861  & - &  2 &- &2&1.5                      & \Hb 4861 & 100 &100 &
100&100 \\
 \Hb 4861$^3$  & - & 1.75  &- &2.5 &2                & \Hb 4861$^4$ &  - & 6.57
& 0.15&0.098   \\
 - &-&-&-&-& -                                  &        [OIII] 5007+&1145 & 1150&
970.&1074. \\
  \Vs (\kms)&-&300& - &  150.& 150.   &         -    &-& 30.&50. & 50.\\
  \n0 (10$^6$ $\rm cm^{-3}$)&-&2 & - & 40. & 40. &         -  & - & 0.15  & 0.15&0.15  \\
  \B0 (10$^{-3}$ gauss)&-& 1&- &  1&  1&    -  & - & 1&1&1 \\
  \Ts (10$^5$ K) &-& 1.04 & - &0.6&  0.6&       -  & - & 1.04& 0.6&0.6 \\
   U      &-& 10 & -& 4.5& 4.&                 -  & - & 0.007&0.013&0.08 \\
  $d/g$ (10$^{-4}$)& -& 12& - &4& 0.2&              -  &- &  20    & 4 &0.04   \\
  \agr (\mum)    &-&0.2&-&0.2&2.&   -   &      -& 0.2&0.2&2. \\
  C/H (10$^{-4}$) & - & 5.3 & - & 5.3& 5.3&      -  &  - & 3.3 &3.3&3.3  \\
  N/H (10$^{-4}$) & - & 5.1 &-& 5.1&5.1&        - &  - & 0.91 &0.91&0.91\\
  O/H (10$^{-4}$) & - & 7.6 &- & 7.6 &7.6      &- &  - & 6.6 &6.6&6.6\\
  \hline
\end{tabular}
}
\flushleft
$^1$ fluxes in 10$^{-12}$ \erg (Lutz 1984);
$^2$ fluxes in 10$^{-12}$ \erg (Schmid \& Nussbaumer 1993);
$^3$ absolute fluxes in 10$^5$ \erg;
$^4$ absolute fluxes in \erg;
$^5$ blended with [SII].
\end{table*}
\subsection{The UV lines}
The data come from L84 and from SN93.
Notice that the two spectra in the UV (left side of Table \ref{tab:tbl1}) observed in
different times
show   compatible  line ratios, considering that the Lutz (L84) spectrum is  reddening
corrected
while  Schmid \& Nussbaumer (SN93) give the observed fluxes.
However, the NV 1239 line  which  was  observed by Lutz,
is absent in the  SN93 spectrum.
This is a crucial line,  which can be explained not only by a higher \Ts,
but  also by a relatively high shock velocity.
Indeed, the shock velocity is responsible for the
heating of the gas downstream in the immediate post shock region where
T$\propto$ \Vs $^2$.

We have  calculated the spectra by  different models, m1,  m2, and m3 (Table \ref{tab:tbl1}, bottom).
Model m1 leads to the best fit of the calculated  line ratios  to those observed by
Lutz and is characterised by a high \Vs \, (300 \kms) and a high \Ts \, (100,000 K).
Model m2 explains the line ratios observed by Schmid \& Nussbaumer,
who derived a temperature of the hot star    \Ts= 60,000 K.
Such a relatively low temperature is valid to explain the UV spectra which do
not show lines from
relatively high ionization levels (e.g. NV).
Model m3  is characterised by a large \agr,  which is consistent with
crystalline grain formation  (see $\S$4 and 5).
The relatively high magnetic field adopted (\B0 = 10$^{-3}$ gauss) is characteristic of
SS (e.g. Crocker et al. 2001).
Notice that  changing one input parameter implies the readjustment of all
the other ones.\\
The models which explain the UV line ratios correspond to different temperature of the hot star and
different grain radius. \Ts= 100,000 K and \Vs=300 \kms which are used in model m1 to explain the UV
spectrum observed by Lutz, particularly the NV 1240 line flux, are less suitable because such high
velocities are not seen in the FWHM profiles. Moreover, \Ts=100,000 K leads to HeII4686/H$\beta$
higher by a factor of $\sim$ 3 than observed in the optical domain.
Adopting \Ts=60,000 K both the UV line spectra (SN93) and the optical ones (L84 - see $\S$3.2) are satisfactorily explained.\\
 Higher preshock densities are adopted by
models m2 and m3 to compensate for the lower compression downstream
which results from  a lower \Vs \, (=150 \kms).\\
Notice that the shock velocity is higher in the reverse shock (\Vs=150\kms) than in expanding shock (\Vs=50\kms).
The velocity of the reverse shock is rather high (\Vs= 150-300 \kms)
compared with
the ones obtained from the radial velocity measurements by L84 (Table 5 therein),
which are $\sim$ 12-16 \kms.
The densities in the reverse shock
are too high to give a  contribution to the optical forbidden lines (e.g. [OII]).\\
At the bottom of Table \ref{tab:tbl1} the model input parameters are shown.
The relative abundances  C/H, N/H, and O/H appear in the last rows.
The other elements (H, He, Ne, Mg, Si, S, Ar, Fe)  are adopted with solar
abundance
(Allen 1973) because no lines of these elements are available.
Indeed, P05 indicate a near solar Fe/H.
The  relative abundances  adopted  for models m1,  m2, and m3, consistently calculated, are C/O=0.70, N/O=0.67, and C/N=1.04 in the reverse shock, while
L84   found   C/O=0.79 and N/O= 1.00
and  SN93 C/O=0.73, N/O=0.67, and C/N=1.1. The relative abundances of C, N, and O adopted to reproduce the UV spectra are in good agreement with those obtained by SN93 for symbiotic stars. Moreover, the models lead to CIII]1909/SiIII]1892 $<$1, which indicates that HD330036 is less adapted to the PN class.
\subsection{The optical lines}
The optical spectra observed by Lutz (Table \ref{tab:tbl1}, right) contain several forbidden lines which refer to relatively low critical densities for
deexcitation, particularly  the [OII] lines. The radial velocities
observed by Lutz are rather low and applicable to those found in the
winds close to the giants,  typical of 20-30 \kms.
 The densities  tend to decrease with distance from the giant centre: therefore the outward shocks most likely  accompany
the   wind close to the system.

Lutz found that the emission line region is characterised by electron densities $n_e\sim$ 10$^6$ \rm cm$^{-3}$ and temperatures
of \Te $\sim$ 1.5 10$^4$ K. Actually, the weakness of the [OII]
lines compared to the [OIII] lines and the strong [OIII] 4363 line
are indicative of relatively  high \Te\ and $n_e$. The presence of
strong optical forbidden lines constrains the density to $\leq$
10$^6$ $\rm cm^{-3}$. Recall that the densities downstream are
higher than the pre-shock ones by about a factor of 10 due to
compression. The agreement of the calculated optical  line ratios
with the data, adopting solar abundances, indicates  that  the shock
is sweeping up the IS matter. Moreover the models lead to grain
temperatures $<$ 100 K.

The results of model calculations reproduce the data within the errors
(20-30 \%),  except the observed HeII 4686 line in model m4, which is overpredicted
by a factor of $\geq$ 3: therefore this model is less reliable.

\begin{figure*}[!hp]
\begin{center}
\includegraphics[width=0.4\textwidth]{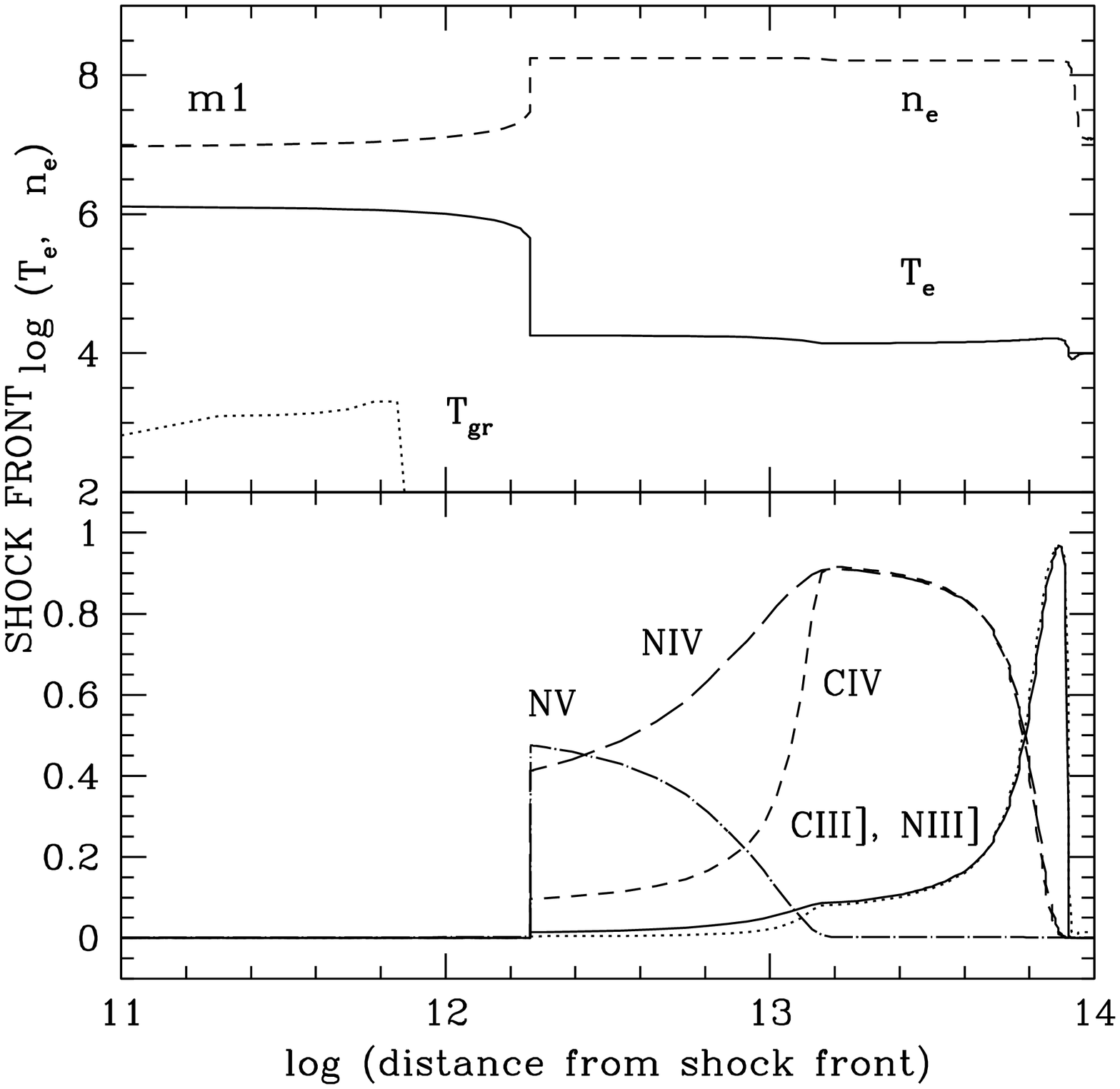}
\includegraphics[width=0.4\textwidth]{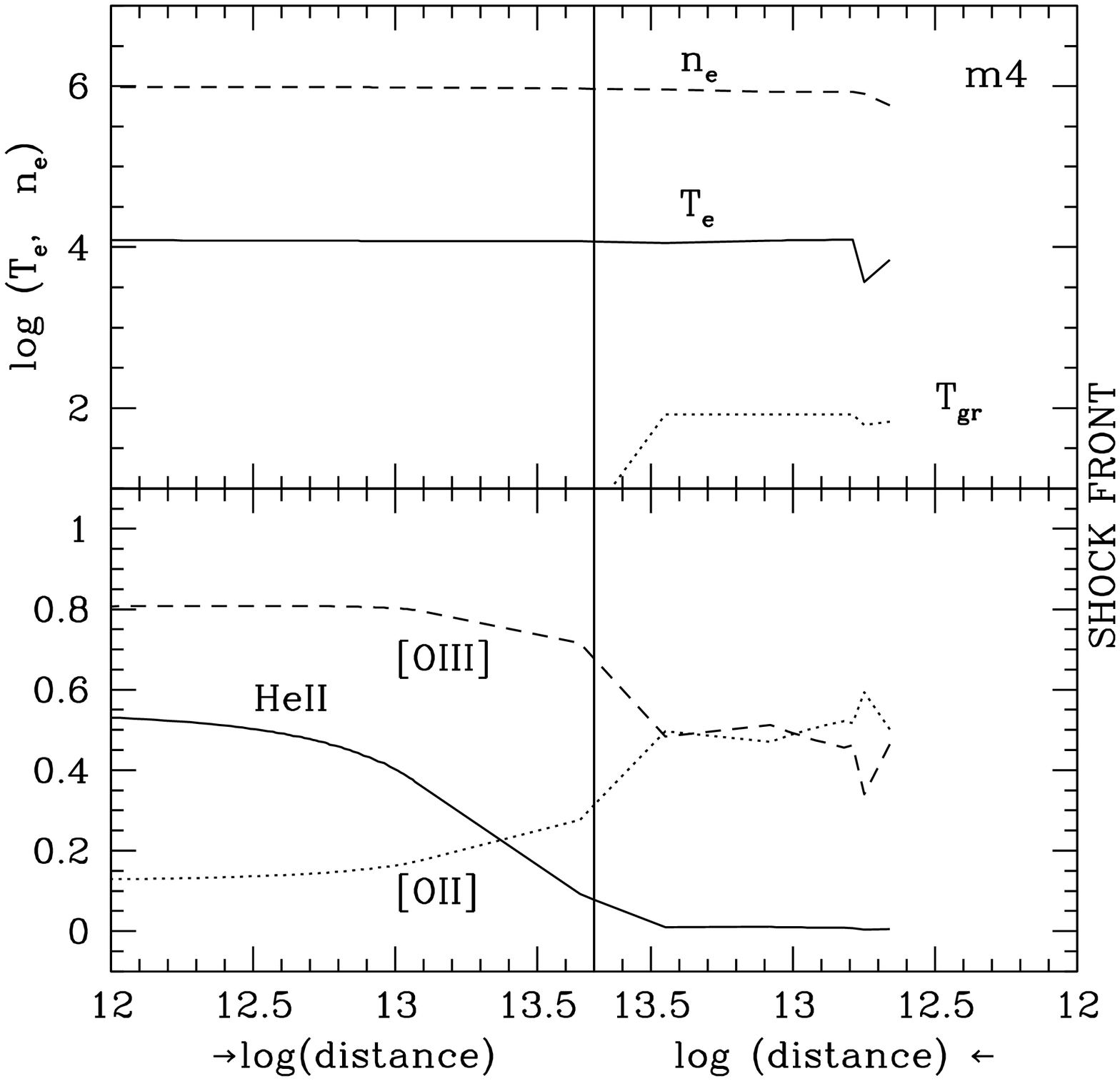}
\includegraphics[width=0.4\textwidth]{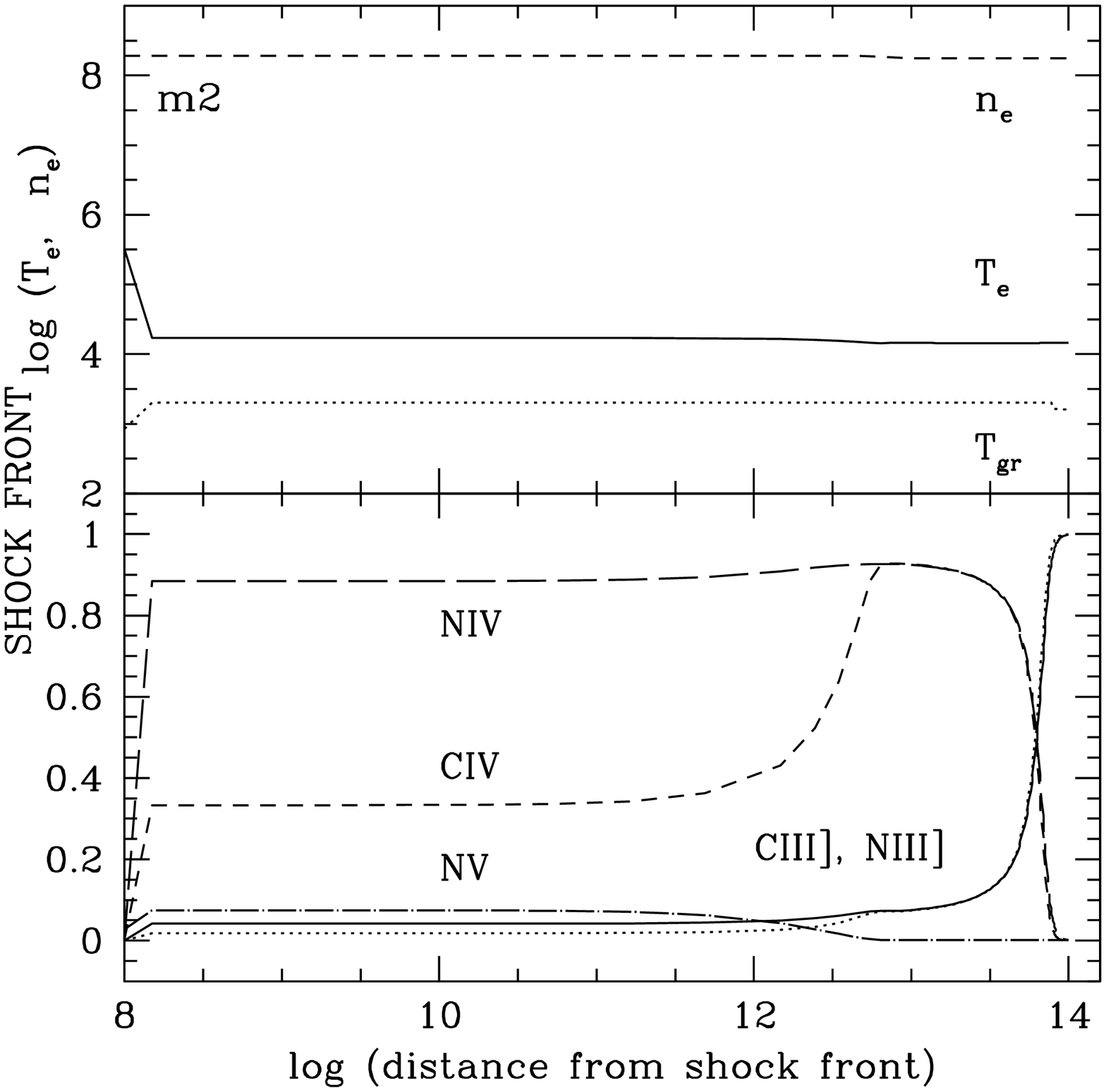}
\includegraphics[width=0.4\textwidth]{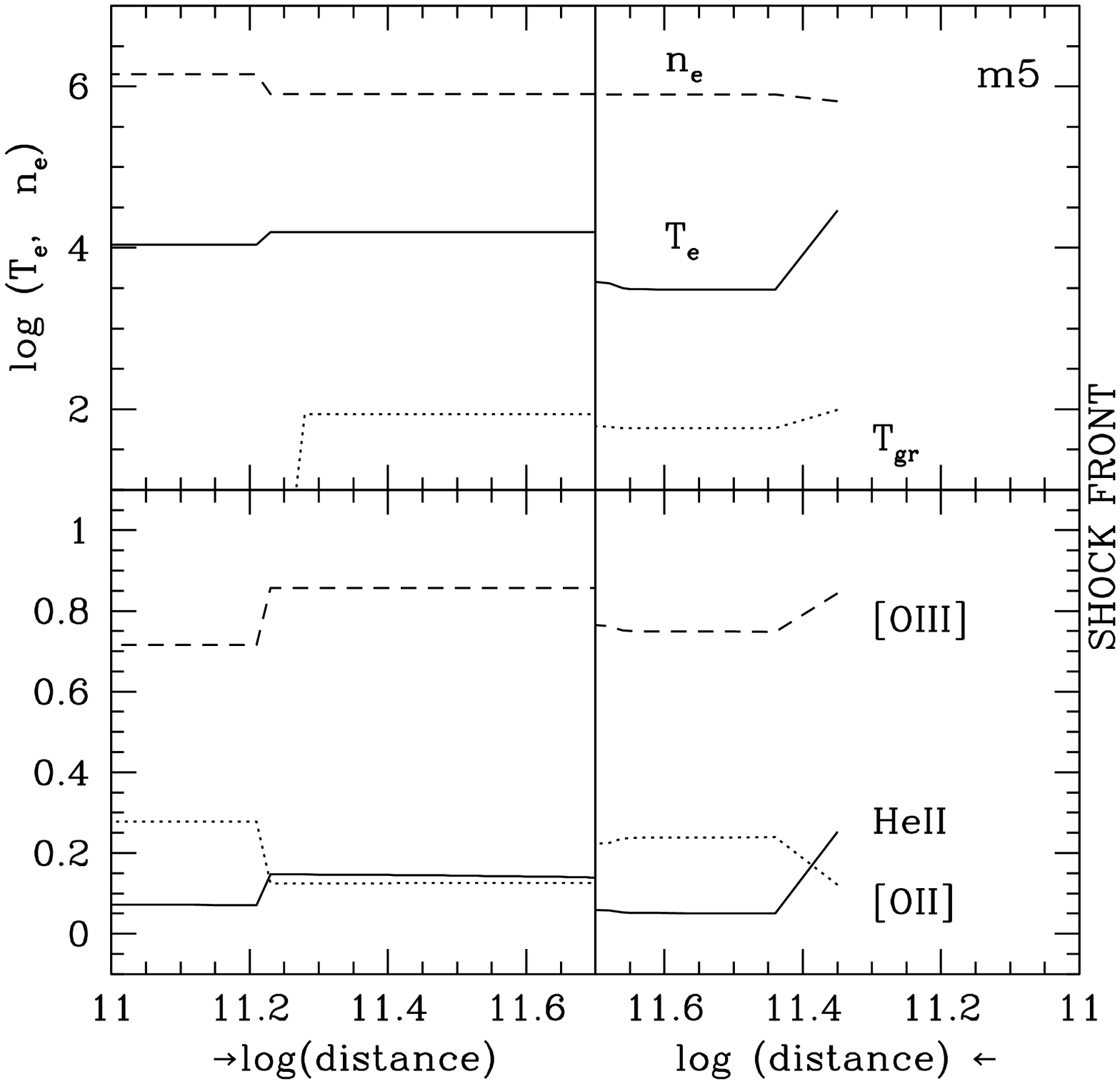}
\includegraphics[width=0.4\textwidth]{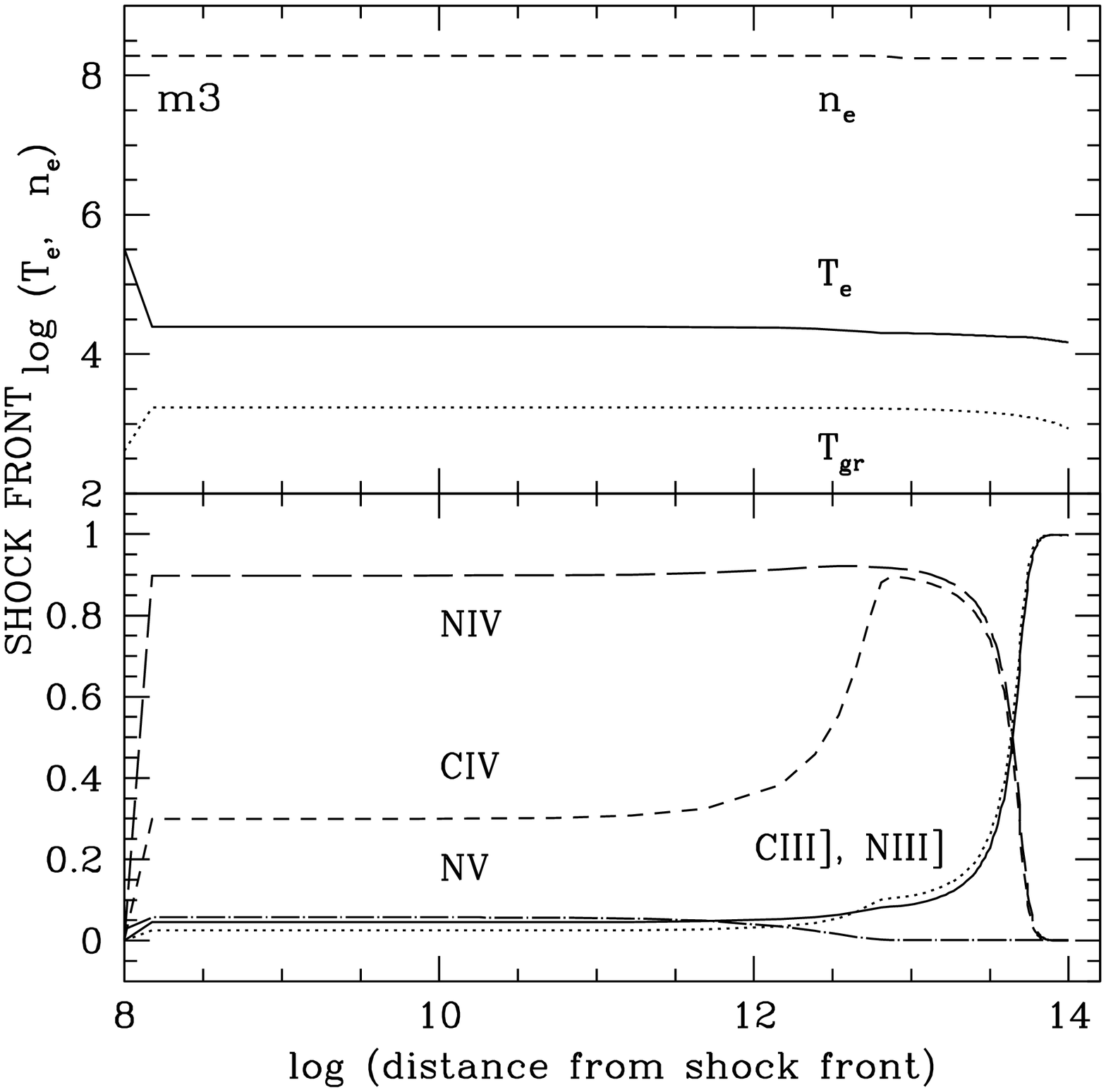}
\includegraphics[width=0.4\textwidth]{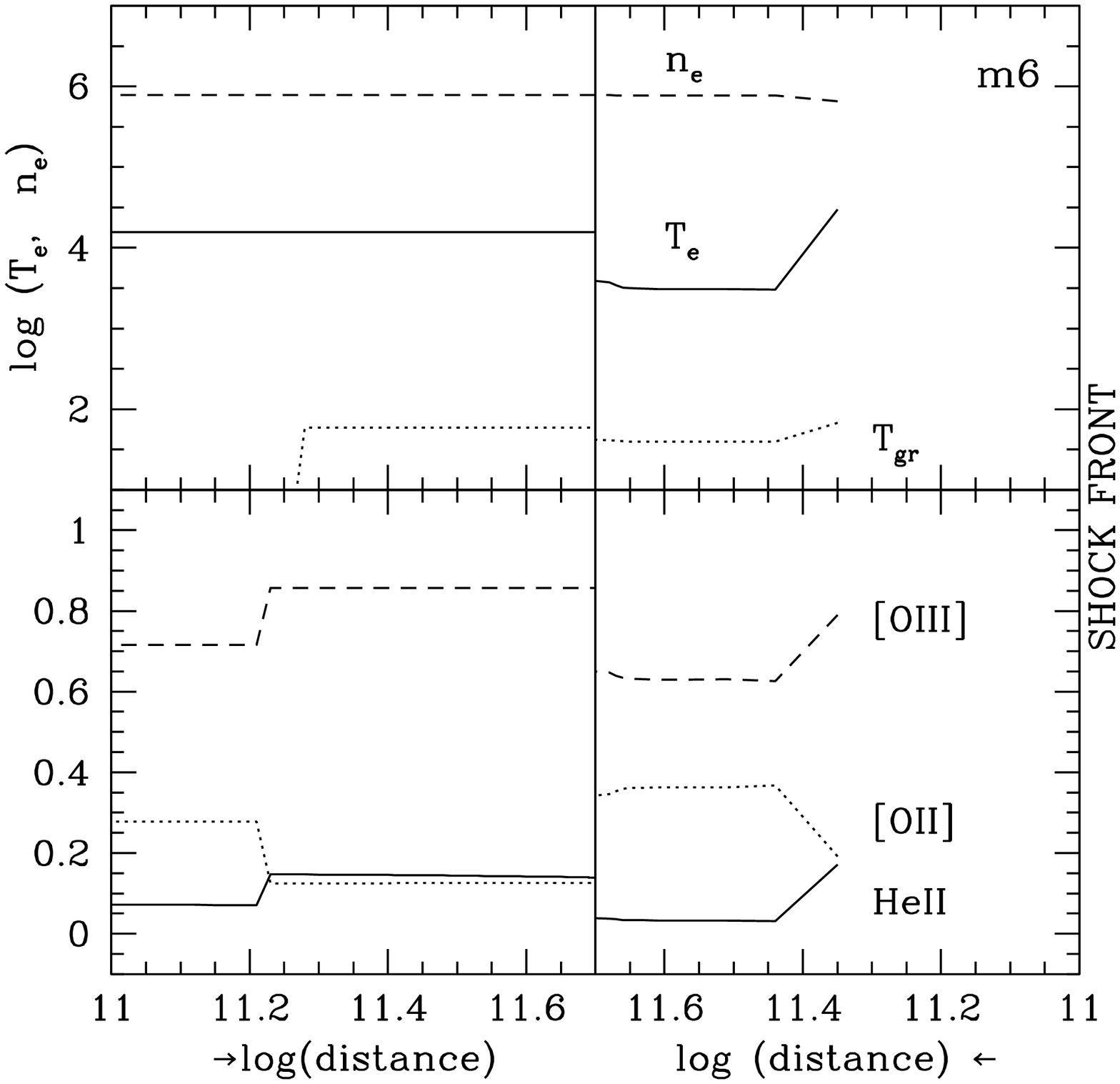}
\caption{Top left diagram. Top panel: the profile of the electron temperature
(solid line),
of the grain temperature (dotted line), and of the electron density (dashed line)
downstream  of the reverse shock for model m1;
bottom panel: the distribution of the fractional abundance downstream.
Top right diagram: the same  downstream of the expanding shock for model m4.
Middle left diagram: the same  for model m2.
Middle right  diagram: the same  for model m5.
Bottom left diagram: the same  for model m3.
Bottom right diagram: the same  for model m6.\label{fig:tt}}
\end{center}
\end{figure*}

In Fig. \ref{fig:tt} the profiles of the electron temperature and density
downstream of the reverse shock (left) and of the expanding shock
(right) are given as well as the distribution of the fractional
abundance of the most significant ions (bottom panels). The
photoionizing source (hot star) is on the left for all diagrams:
therefore in the diagrams  on the right of Fig. \ref{fig:tt}, which refer to
the case in which the photoionizing flux and the shock act on
opposite edges, the distance scale is logarithmic and symmetrical
with respect to the centre of the nebula with the highest resolution
towards the edges of the nebula. In the top diagrams models m1
(left) and model m4 (right) appear: notice that in the nebula
downstream of the reverse shock represented by model m1, sputtering
is very strong and  the grains are destroyed after a small distance
from the shock front. In the middle diagrams the physical conditions
for models m2 (left) and m5 (right) are described, while  bottom
diagrams  refer to models m3 (left) and m6 (right). The comparison
between the middle and bottom diagrams shows that relatively large grains
(\agr = 2 \mum) affect the distribution of  the physical conditions
downstream, particularly, the  distribution of the ions. \\
The $d/g$ ratios is selected cross-checking the modelling of the
continuum: a high $d/g$ enhances the cooling rate
downstream, changes the distribution of the physical conditions and, consequently, the emitted line intensities.
\section{The continuum SED}
\begin{figure*}[!ht]
\begin{center}
\includegraphics[width=0.47\textwidth]{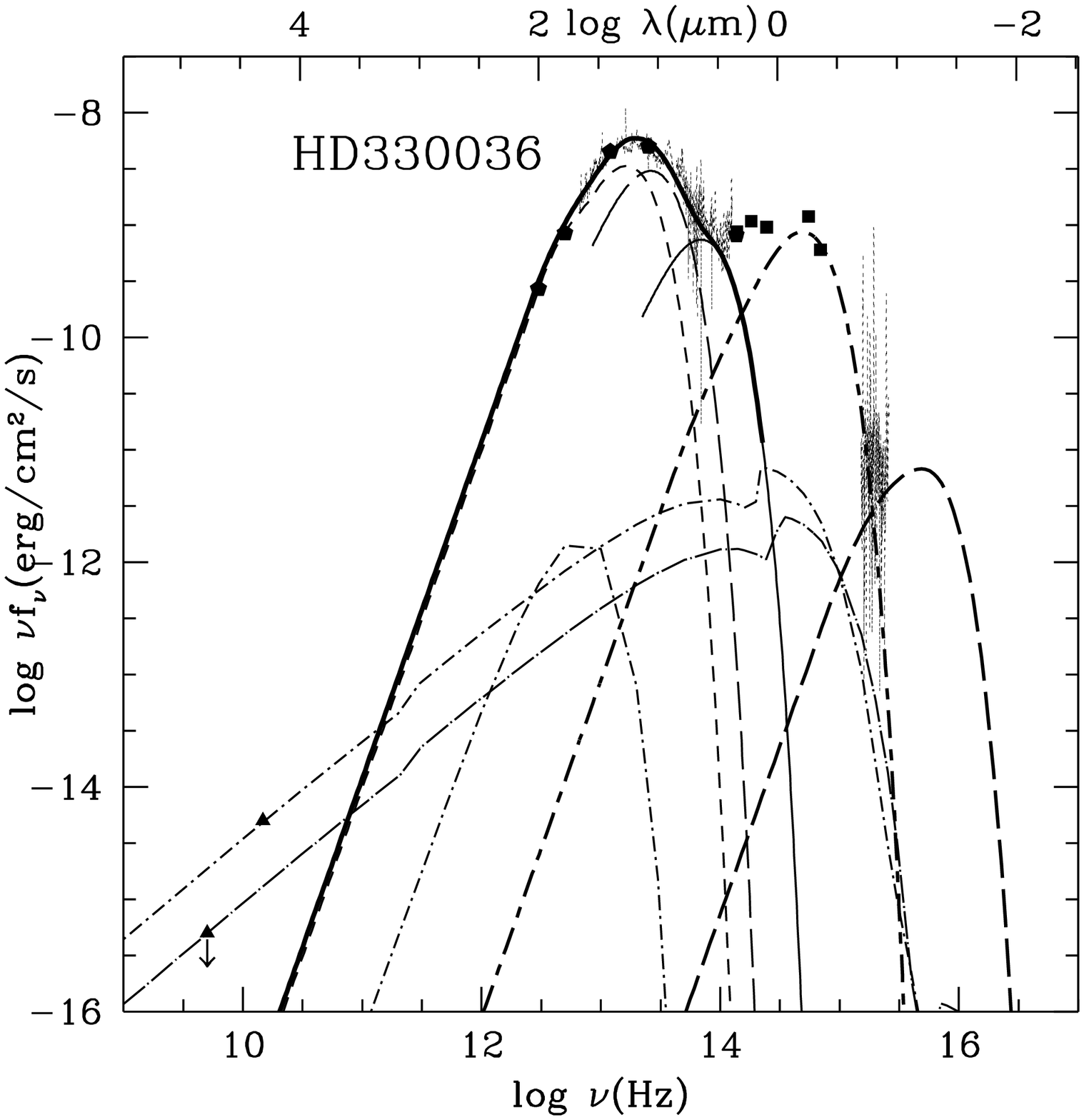}
\includegraphics[width=0.47\textwidth]{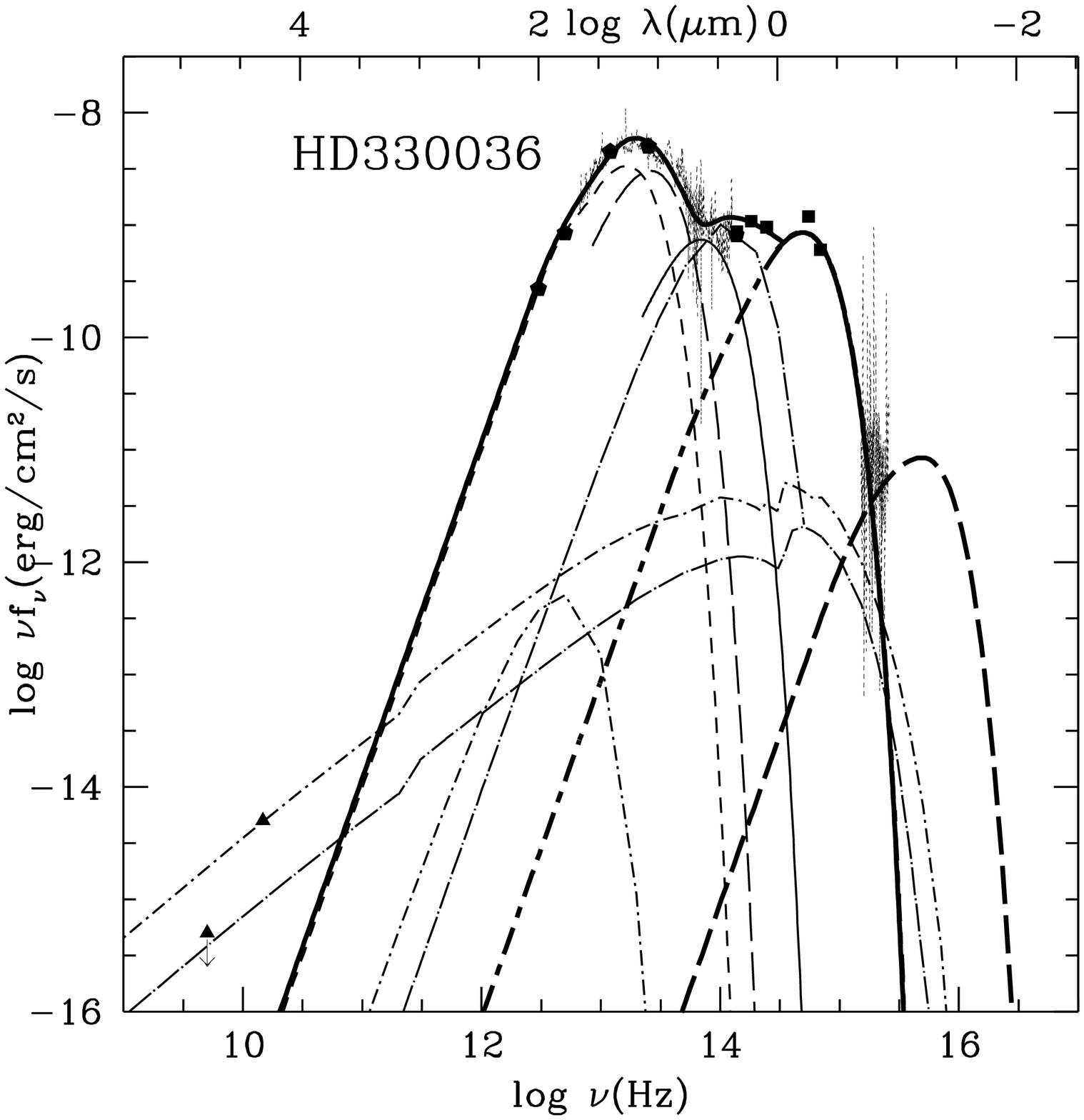}
\caption{The SED of the continuum. Data (black squares) from the IUE archive
(UV range), from ISO archive, IRAS catalogue of Point Sources and 2MASS
database (IR range), from Kharchenko 2001 (B and V Johnson magnitude)
and from Milne \& Aller 1982 (black triangles - radio range).
The bb at  60,000K (thick long-dashed); the bb at 6000K (thick long-short dashed); the bb at  850K (thin solid}; the bb at  320K  (thin long-dashed); the bb at 200K (thin short-dashed). Left diagram: the bremsstrahlung  calculated by model m2 (long-dash dot); the bremsstrahlung and relative dust emission  by model m5 (short-dash dot); the summed SEDs of the dust shells (thick solid). Right diagram: the bremsstrahlung and relative dust emission  calculated by model m3 (long-dash dot); the bremsstrahlung and relative dust emission  by model m6 (short-dash dot); the summed SEDs of the dust shells, reemission by dust from the reverse shock, and the cool star bb (thick solid).\label{fig:seds}
\end{center}
\end{figure*}
In Fig. \ref{fig:seds} we present the modelling of the continuum SED.
The data in the IR come from the ISO archive (see $\S$5 for more
details), the IRAS Catalogue of Point Sources and the 2MASS database, while
the data in the UV are extracted from the IUE archive and refer to the
1984 spectrum already analysed by SN93. The radio points are from
Milne \& Aller (1982) and the optical ones (B and V Johnson magnitude)
from Kharchenko (2001).

The SED of the continuum  results from the contributions of the  
fluxes from the cool and hot stars, as well as the fluxes from the dust shells, the
bremsstrahlung from the ionized nebulae downstream of the shock
fronts (which emit the UV and optical line spectra) and the reprocessed
radiation from dust.

At a first guess, the flux from the stars and the dust shells
is approximated by black body radiation.
 We find that  a black body of 6000 K fits
the NIR data corresponding to the cool star, in agreement with P05,
while the temperature of the hot star is calculated
phenomenologically by modelling the line spectra (\S3).
\subsection{The shells}
The dust grains form in the atmosphere of the giant star where the temperature has dropped below $\sim$ 2000 K and the
densities are $\sim$ 10$^9$ - 10$^{10}$ $\rm cm^{-3}$. 
Their temperature results from the
coupled effect of the  photoionization flux from the WD and collisional
heating by the gas. The
flux emitted from the region closer to the shock front downstream,
which corresponds to the maximum temperature of dust, is calculated by the
Plank-averaged absorption coefficient of dust. The cooling rate is
very strong downstream at such densities, therefore the flux from
the shell corresponds mainly to the maximum temperature of the grains and 
can be modelled by a black body.
The temperature of the dust shells is derived  by modelling the
ISO and IRAS data which agree in the overlapping frequency range: this suggests that variability, at least in the IR domain, has not been as so large as to substantially modify the dust continuum emission.
Fig. \ref{fig:seds} diagrams  show that the  data can be explained by the
combination of at least three black body (bb) fluxes,  corresponding to
temperatures of 850 K, 320 K, and 200 K.

By comparing the  models with the  data we obtain the $\eta$
factors: they depend on the radius of the dust shell, $r$, and on the 
distance of the system to Earth $d$ ~ ($\eta=r^2/d^2$), being the fluxes calculated at the nebula and the data measured at Earth.
Adopting \textit{d}=2.3 kpc, we find \textit{r}= 2.8 10$^{13}$ cm, 
4 10$^{14}$cm, and 10$^{15}$ cm
for the shells at 850 K, 320 K, and 200 K, respectively.
Interestingly, this implies that all the dust shells are circumbinary, with the
coolest ones extending well beyond the two stars if we assume an upper limit for
binary separation of $\sim$ 8 10$^{12}$ cm (5 R$_g$), as suggested by Z06.
According to the D' type nature of this star, the dust shell at $\sim$ 1000 K generally observed in D type SS does not appear in HD330036.
\subsection{The shocked nebul\ae{}}
The radiation emitted from the shocked nebulae accounts for both bremsstrahlung and
dust reprocessed radiation, which are calculated consistently in  the downstream
region. The  fluxes  are integrated from the shock front  throughout regions of gas and dust in  different physical conditions. The reradiation  IR bump from dust  is in fact wide because it accounts for the stratification of the temperature downstream.
The bremsstrahlung covers a large wavelength range, from radio to  UV.

In the previous sections we have presented some alternative models which were selected
from the best fit of the line spectra.  Since the models m1 and m4, calculated adopting a hot star temperature of \Ts $\sim$ 10$^5$ K, fail in reproducing the HeII 4686/\Hb~ ratio, in the following we consider only the remaining ones (Table \ref{tab:tbl1}). The continua calculated by these models are compared with the observations in the diagrams of Fig. \ref{fig:seds}: in the left one the models m2 and m5, representing the nebulae downstream of the  reverse and of the expanding shock respectively, and calculated with a hot star temperature of \Ts=60,000 K and an initial grain radius \agr=0.2 \mum, appear; in the right one, models m3 and m6, calculated with \Ts=60,000 K and \agr=2 \mum, are shown. Models with \mum-size grains are justified by the presence of silicate crystalline features (e.g. Vandenbussche et al. (2004)) at the top of the IR continuum, as we will discuss in details in $\S$5.3. 
In Fig. \ref{fig:seds} diagrams we have added also the UV data in order to  constrain
the bremsstrahlung fluxes in the UV range, while the modelling of the IR
data is  constrained by the $d/g$ ratios. \\
The calculated line ratios and the continuum have been then cross-checked until both the line spectra and the SEDs were fitted satisfactorily.

The SED of the bremsstrahlung  is constrained by the radio data,
and the dust reradiation peak by the IR data in the 1-3 \mum\ range.
Generally in D-type SS the cool star is of Mira type with temperatures
of 2000-3000 K, therefore their fluxes cover the data in the NIR range.
However HD330036 shows a cool star temperature of 6000 K, therefore adopting the bb
approximation, the emission peak results shifted toward higher frequencies and the
1-3 \mum\ continuum data are most probably explained by hot dust.

The right diagram of Fig. \ref{fig:seds} shows that the contribution of the reprocessed
radiation of dust from the shocked nebula downstream of the reverse shock
(m3) is significant in this range, while the grains downstream of the
reverse shock calculated with \agr=0.2 \mum\ (m2) reach temperatures
of $\sim$ 1900 K and easily evaporate: therefore model m2 is not significant in the modelling of the hot dust. The contribution of the dust downstream of the expanding shock which reaches temperatures of $\sim$ 100K cannot be seen in the SED because hidden by the dust shell radiation flux.

The $d/g$ ratio  for models m2 and m5 is 4 10$^{-4}$, of the order of $d/g$ in the
ISM; for models m3 and m6 the $d/g$ ratio is even lower, being reduced by factors $>$ 10.\\
Recall that dust  emissivity at a temperature T$_d$ is
calculated by 4$\pi$ a$_{gr}^2$ B($\lambda$,T$_d$) $d/g$ n$_{gas}$ (Viegas \& Contini 1994), where B($\lambda$,T$_d$) is the Plank function. A lower d/g, which is constrained by the data, compensate for a higher \agr. Therefore, in the two diagrams
of Fig. \ref{fig:seds} which account for models calculated by different \agr, the intensities of the reradiation peak of dust relative to  bremsstrahlung are similar.

We can now calculate the radius of the nebulae by the $\eta$ factors.
Adopting d= 2.3 kpc and models m2 and m5, the reverse shock and the
expanding shock have r=1.9 10$^{13}$ cm and r=4.9 10$^{15}$cm
respectively, while adopting models m3 and m6 the
reverse and expanding shocks have radius r=1.8 10$^{13}$ cm
and r=8.7 10$^{16}$ cm, respectively. \\
It is worth noticing that the reverse shock radius is an upper
limit because we have adopted the maximum value for the distance (d=2.3 Kpc) and the $\eta$ is constrained by the datum at 5 GHz, which in turn is an upper limit.
Fig. \ref{fig:tt} (bottom left) shows that the
temperature of dust calculated with model m3 is  $\leq$ 1500 K  at a
distance  $>$ 10$^{12}$ cm from the shock front, in agreement with a
shell radius  of $\sim$ 10$^{13}$ cm calculated from the  $\eta$
factor by fitting the SED in Fig. \ref{fig:seds} (right diagram). 

Finally, the datum at 14.7 GHz in the radio range  constrains the bremsstrahlung, 
whereas the other one at 5 GHz is just an upper limit (Milne \& Aller 1982).
Incidentally, the physical conditions  downstream  of models m1, m2, and m3, representing the reverse shock, lead to  an optical thickness $\tau$ $>$ 1 at $\nu$ $<$ 10$^{12}$ Hz (Osterbrock 1988), indicating that self absorption of free-free radiation  reduces the flux. On the other hand, the conditions downstream of models m4, m5, and m6,
which represent the expanding shock, lead to $\tau$ $<$ 1 at 14.7 GHz but  to $\tau$ $\geq$ 1 at 5 GHz.

Summarising, we have found that the physical parameters which best explain the
shocked nebulae are \Ts=60000K, \Vs=150 \kms, \n0=4 10$^{7}$ \cm3, \agr=0.2 \mum \, for the reverse shock, while for the expanding shock we found \Vs=50\kms, \n0=1.5 10$^5$ \cm3 and grains of both sizes,\agr=0.2 \mum \, and 2 \mum.
\begin{figure*}[!ht]
\begin{center}
\includegraphics[width=0.47\textwidth]{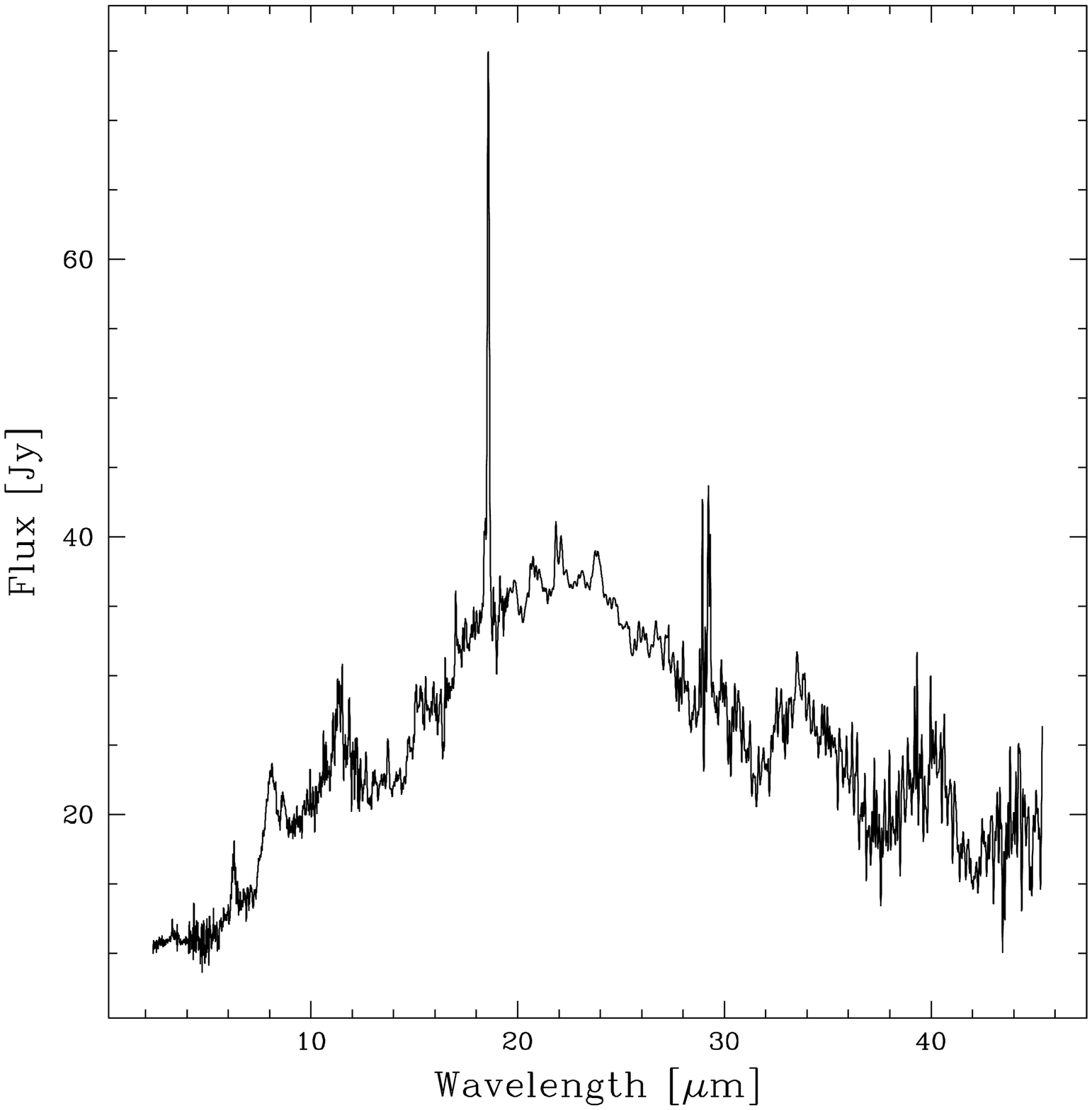}
\includegraphics[width=0.47\textwidth]{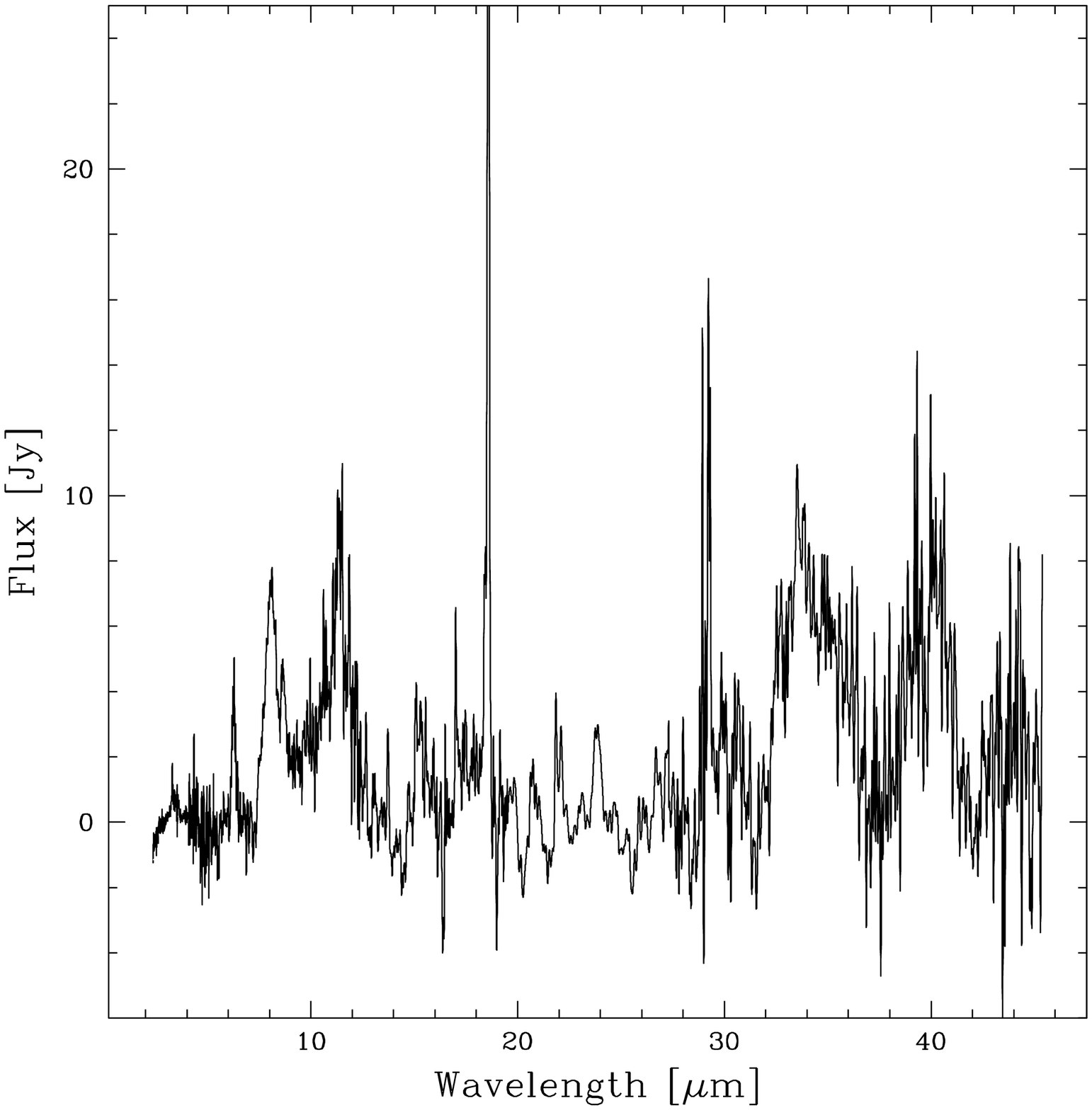}
\caption{Left: ISO-SWS spectrum of HD330036. The strongest spectral features at $\sim$ 18 \mum\ and $\sim$ 28 \mum\ are instrumental artifacts (see \S5.1). Right: continuum subtracted ISO-SWS spectrum of HD330036. Notice the bands usually attributed to Polycyclic Aromatic Hydrocarbons (at 3.3, 6.2, 8 and 11.3 \mum). At longer wavelengths we found evidence for emission from crystalline silicates
(clearly visible the strong complexes at $\sim33$ \mum \, and $\sim40$ \mum)
\label{fig:sws}}
\end{center}
\end{figure*}
\section{The solid state features in the ISO spectrum}
\subsection{Observations and Data Reduction}
HD330036 was observed by ISO on 1996, September 18. \\
In this paper we focus on the spectrum taken with the Short
Wavelength Spectrograph (hereafter SWS - de Graauw et al. 1996),
used in full-grating scan mode (AOT 01) and covering the wavelength
range between 2.38 and 45.2 \mum \, (Fig. \ref{fig:sws}, left). The
spectrum comes from the ''Uniform database of SWS 2.4-45.4 micron
spectra'' within the Highly Processed Data Products (HPDP) section
of the ISO Archive. This database presents a complete set of all
valid SWS full-scan 2.4-45.4 \mum\ spectra processed (from the last
stage of the pipeline software) and renormalised in as uniform a
manner as possible: in particular, the processing produces a single
spectrum for each observation from the 288 individual spectral
segments, which are the most processed form available from the ISO
archive (see Sloan et al. 2003 for details about the algorithm used
to generate the database). However some instrumental artifacts are
still present (e.g. the strong ''glitches'' at $\sim$ 18 \mum \, and $\sim$ 28 \mum, whose
profiles show a characteristic vertical flux increase followed
by an exponential decrease, revealing then their fake origin): obviously, these
features are not considered in deriving physical conclusions and we point them out whenever the real origin of such structures remains ambiguous.

We have analysed the data with the IRAF package software. In particular, we have
defined a continuum for the ISO-SWS spectrum by means of a spline-fit
function: although this continuum has not necessarily a physical meaning, it merely
aims to enhance the sharp structures visible on top of the spectrum and to offer
an easier identification of the solid state features.
The continuum subtracted spectrum of HD330036  appears in Fig. \ref{fig:sws}, right.
In spite of the low quality, particularly at long wavelengths,
many dust bands are recognised.

The spectrum appears substantially different with respect with the other SS ISO spectra analysed by Angeloni et al. (2007a), where the mid-infrared range is dominated by the strong emission of amorphous silicate dust. Conversely, in HD330036 it can be clearly split into two regions: the short wavelenght one (up to $\sim$15 \mum) dominated by  PAH prominent emission bands and the long one showing a blending of narrow and characteristic emission profiles, often concentrated in specific wavelength regions (named \textit{complexes}),
whose carriers are believed to be crystalline silicates.
Only after the ISO mission these  dust species were noticed to be present
outside our own solar system, allowing for the first time a mineralogical analysis of the dust composition in a wide range of
astrophysical environments. In the following, we study the specific solid state features visible in the ISO-SWS spectrum. 
\subsection{PAHs}
A wide variety of astronomical sources show
some strong emission bands  at 3.3, 6.2, 7.7, 8.6 and 11.3 \mum:
the so called unidentified infrared (UIR) emission features
(Puget \& Leger 1989, Allamandola et al. 1989). Though some features remain
still unidentified,
the UIR spectrum as a whole is  linked to PAHs,
or to small grains containing PAHs, whose variety and complexity
 suggest a mixture of ionized and neutral PAH molecules.\\
HD330036 exhibits all these UIR bands (Fig. \ref{fig:hdpah}).
In this section we analyse
their profiles and peak positions, in order to investigate the PAH local conditions and
formation history.
\subsubsection{The 3.3 \mum \, feature}
HD330036 shows a prominent feature at $\sim 3.29$ \mum \,  (Fig. \ref{fig:hdpah}a)
usually attributed to the C-H stretching mode \mbox{($v= 1\rightarrow0$)} of highly excited PAHs.
The profile is clearly asymmetric, with the blue wing steeper than the red one:
the peak position and the FWHM (3.292 and 0.037 \mum, respectively) look similar to other 3.3 \mum \, profiles seen in e.g. the  Orion bar, even though these characteristics  seem to be not so common in astronomical sources (van Diedenhoven et al. 2004).\\ A second weaker feature appears centred at $\sim 3.40$ \mum ~and is identified with the excitation of higher vibrational levels (Barker et al. 1987). There are unconvincing proves of evidence for the other, more weaker emission features at 3.44 and 3.49 \mum.\\ According to laboratory studies concerning the role of the physical environment on the IR spectroscopy of PAHs (Joblin et al. 1994, 1995) and on the basis of the band profile, we suggest that the carriers of the 3.3 \mum\ feature in HD330036 are likely to be large PAH molecules, at rather high temperatures ($\sim$ 800-900 K). Although it is far from being conclusive, it is worth noticing the similarity between the observed ''symbiotic'' profile and the laboratory one of the ovalene molecule, as reported by Joblin et al. (1994-1995).
\subsubsection{The 6.2 \mum \, feature}
Even if this region of the spectrum is moderately noisy, we easy recognise the
feature at $\sim 6.2$ \mum \,
(preceded by a weak feature at about 6.0 \mum)
 which is the PAH C-C stretching band (Fig. \ref{fig:hdpah}b).
The precise peak position and the width of this emission feature are strongly
influenced by
several parameters,
e.g.  molecular size, molecular symmetry, charge status, dehydrogenation, etc.
(Hudgins and Allamandola,
1999; Bakes et al. 2001; Hony et al. 2001).\\
The overall shape of the profile, peaking at $\sim 6.25$ \mum \, and rather
symmetric, suggests a
link with objects such as some post-AGB and emission line
 stars. According to  e.g. Peeters et al. (2002) this symmetry could indicate
that the PAH family
emitting the band at $\sim 6.2$ \mum \, has not yet been exposed to a harsh
radiation field
and its composition still reflects the conditions during the formation at high
temperatures.\\
This last remark, along with the temperature suggested by the 3.3
\mum \, band, is consistent with our scenario proposing that
PAHs within HD330036 lie in the inner region (T $\sim$ 850K, r $\sim$
2.8 10$^{13}$ cm) as found by  modelling  the SED ($\S$4).
\subsubsection{The 7.7 and 8.6 \mum \, features}
The ''7.7 \mum'' feature in HD330036 appears clearly redshifted with respect to
standard
positions observed in other astronomical sources (Peeters et al. 2002).
Its profile seems to show several substructures (Fig. \ref{fig:hdpah}c);
  furthermore the peak position is at $\sim 8.08$ \mum, and there is no apparent
trace of
the two main components seen in the ''standard'' profiles at $\sim 7.6$ and
$\sim 7.8$ \mum, respectively.
The band resembles the one seen in the H-deficient Wolf-Rayet star (WR 48a), hence
the whole feature could be a sort of blend of ''classical'' 7.7 \mum \,
 PAH feature and of a UIR band whose carriers seem likely to be amorphous carbon
dust or large ''pure''
carbon molecules (Chiar et al. 2002).
The band usually ascribed to C-H in plane bending vibrations of probably ionized
PAHs
at $\sim 8.64$ \mum \, is also present.
\subsubsection{The 11.3 \mum \, feature}
\begin{figure}[!hb]
\includegraphics[width=0.4\textwidth]{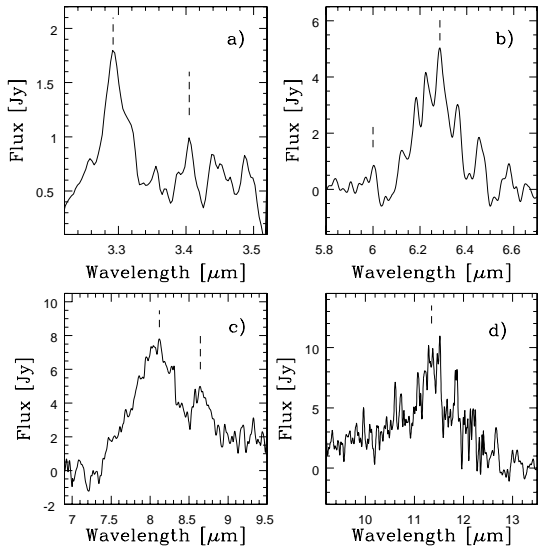}
\caption{The UIR bands in the spectrum of HD330036: a) the 3.3 \mum\, feature;
b) the 6.2 \mum \, feature; c) the 7.7 and 8.6 \mum \,
features; d) the 11.3 \mum \, feature. The dashed lines indicate the band peak
position. \label{fig:hdpah}}
\end{figure}

The strongest PAH band in HD330036 is that at 11.3 \mum, already noticed by Roche et al. (1983).
This range of the spectrum (Fig. \ref{fig:hdpah}d) can show both the bands belonging
to PAHs and to silicates; moreover, unfortunately the S/N level of the detector is severely inadequate: therefore any firm conclusion based on the analysis of the profile is precluded. Nevertheless, some ''peaks'' and a "plateau" do not exclude the presence of some typical, intrinsic substructures.
\subsection{Crystalline silicates}
\begin{figure}
\includegraphics[width=0.4\textwidth]{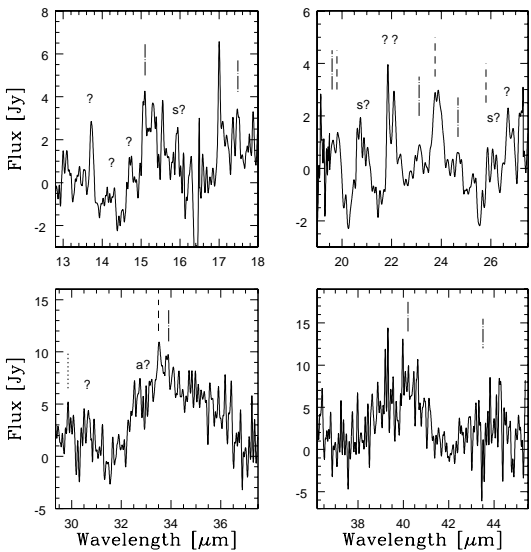}
\caption{Zoom of the spectrum on some interesting crystalline complexes. Short
dashed line: fosterite; dot-dashed line: enstatite; dotted line: diopside; s:
silica; a: anorthite; ?: bands whose attribution to a specific carrier is still doubtful.  \label{fig:hdsc}}
\end{figure}
As stated above, the ISO-SWS spectrum beyond $\sim 15$ \mum \, shows the
presence of bands usually attributed to crystalline silicates.\\
Despite the S/N level not being sufficiently adequate throughout the whole spectral range (e.g. the instrumental band 3E at 27.5 - 29 \mum \, is known for its mediocre performance) several
features are clearly visible on top of the continuum,  constraining the
chemical compositions and spatial distribution of this symbiotic
dust.
\subsubsection{The observed crystalline bands}
A zoom on some interesting crystalline complexes is shown in Fig. \ref{fig:hdsc}. \\
Most of the bands can be confidently identified with crystalline olivine
(i.e. fosterite - $Mg_{2}SiO_{4}$) around 19.7, 23.7 and
33.6 \mum \, and with
pyroxenes (i.e. enstatite - $MgSiO_{3}$) around 23.1, 24.6,
40.5, 43 \mum ,
whereas the features at $\sim$ 15.9, 20.7 and 26.1 \mum ~seem  to agree with
silica ($SiO_{2}$).\\
Several bands still lack an identification (we label them with question marks in Fig. \ref{fig:hdsc}), thus suggesting the presence of more features, even though weaker and noisier.  As already stressed in $\S$5.1, some bands are rather controversial because they could originate from instrumental artifacts (e.g. the 13.8 and 14.2 \mum \, features,
with a contribution from enstatite which cannot be excluded). Furthermore, even when the actual origin of the features has been verified, it is not always easy to attribute them to some specific physical carriers (see, Molster et al. 2002b, Table 1): e.g., those at $\sim$ 20, 26, and 32 \mum \, could fit  the
emission bands
of diopside ($CaMgSi_{2}O_{6}$), as well as  those of anorthite
(i.e. at $\sim $ 26).
However their relative strengths suggest that  they would represent only minor
components.\\
Spectra with a higher S/N ratio, and beyond the wavelength coverage
of ISO-SWS, are  clearly needed to better constrain this
composition insight.
\subsubsection{A disk-like geometry for the silicate dust shell}
After the end of the ISO mission several studies pointed out that
the high abundance of crystalline silicate seems to be related to
the geometry of dust shells. In particular, the objects displaying
strong crystalline bands (e.g. the olivine 33.6 \mum \, band
stronger than 20\% over the continuum) are known to have highly flattened dust
geometrical distributions (Molster et al. 1999a, 2002). The analysis
of the crystalline silicate  profiles in HD330036 revealed that not only the olivine 33.6 \mum \, band is stronger than 30\% over  the continuum, but its profile (with a secondary peak at
34.1 \mum), along with the relative strength of the 23.0 and 23.7
\mum \, features, firmly suggest a disk-like geometry of the silicate envelope. 

The existence of such a disk is also sustained by dynamical
considerations of the orbital parameters of HD330036 as a binary
system: in fact, as noticed by Yamamura et al. (2000) for
close-binary systems like ours, the massive interaction between the
giant and the secondary star strongly affect the local
environment, leading to the formation of a massive circumbinary disk in a rather short time. 
Finally, recall that a disk structure was already proposed for
another D' SS (V417 Cen) whose ring nebula has been optically
resolved (Van Winckel et al. 1994).
\subsubsection{Constraints on dust temperature}
All crystalline silicates in evolved stars tend to show low
temperatures ($<$300K; Molster et al. 2002b). We then suggest that
the crystalline dust temperature in HD330036  lies close to 100-200K.
Otherwise, in stark contrast to the case of higher temperatures (Molster et al. 2002) the strength ratio of the fosterite band at 33.6 and 23.7 \mum \, would have been $<$ 1. This constraint on the temperature, along with the disk
geometry deduced above, indicates that the crystalline silicates
reside in the outer large circumbinary envelopes of dust which were found out by
modelling the IR SED in $\S$4.
\subsubsection{Constraints on the mass loss rate of the cool star}
As reported in the Introduction, an estimate of the mass loss rate of the cool
component of HD330036 is not trivial. As the relations involving stellar atmosphere modelling
are not still reliable (Achmad et al. 1997), we endeavour a different method, by exploiting once again the ISO capability along with the presence of crystalline silicate complexes.  As a matter of fact, the detection limit of ISO for broad spectral features
is - depending on the quality of the spectrum - around 5\% of the continuum
level. The  fact that we clearly recognise several intense crystalline profiles
defines a lower limit for the crystallinity (defined as the ratio between the total
mass of crystalline silicates over the silicate total mass)
and, in turn, an acceptable approximate range for the mass loss rate (Kemper et
al. 2001). In particular, the  ratio of the $\sim$33 \mum \, fosterite band over the
continuum ($\sim$0.4) allows us to suggest that the fosterite mass fraction is
greater than 0.15 and that the cool component of the  symbiotic star HD330036
loses mass at a rate of $\sim$0.4-1 $10^{-6} M_{\odot}/yr$ (see Fig.7 of Kemper et
al. 2001). Unfortunately the $\sim$43 \mum \, enstatite band is located
just to the upper edge of the SWS spectrum, and its exact profile is more ambiguous:
nonetheless, a raw estimate of its relative intensity seems to support our
evaluation of the HD330036 cool component mass loss.
\subsubsection{Crystallization processes}
The crystalline grains require a high temperature  and a low cooling
rate for annealing. Molster et al. (1999a) suggested  that in the
outflow sources the crystallization of silicates takes place close
to the star, i.e.  at high  temperatures,  whereas in a long-term
stable disk crystallization occurs slowly at low temperatures under
the influence of UV radiation
(Sylvester et al. 1999). The temperature required  for an effective  annealing ($\sim$1000K)
which would last long enough to activate the crystallization process
can be reached both in the inner region of the outflowing envelope
of cool component stars  and in the expanding shocked nebula  of
SS, where colliding winds enable shock waves to compress and heat
the dust grains. In effect the shocks could be a
very suitable mechanism thanks to the sudden heating and
gradually cooling of the grains in the post-shock region, favouring the
annealing processes and letting start, in this way, the crystallization path. Such a
mechanism has already been suggested with the purpose of explaining the formation of crystalline
grains within comets in the protoplanetary solar nebula (Harker \& Desch 2002) and claims for a non
secondary role of shocks in the dust transformation processes and consequent infrared emission feature. \\
Our models show that at typical expansion velocities of $\sim$ 15-20 \kms, the dust grains formed at T $\sim$ 1000 K would spend 160-320 d at temperatures higher than 900 K: this period is sufficiently long for the annealing and the subsequent crystallization of a significant portion of dust grains. The crystallization process will occur within a distance of $\sim$  10$^{14}$ cm, which agrees with the size of the outer dust shells.

On the basis of the  theoretical Silicate Evolution Index (SEI)
proposed by Hallenbeck et al. (2000) and the insight of Harker \&
Desch (2002), we thus suggest that crystallization processes in
HD330036 are triggered by shocks and annealing takes place within the circumbinary disk.
\section{Discussion and concluding remarks}
The analysis of the D' type symbiotic system HD330036 is presented
by modelling the continuum SED as well as the line and dust spectra within a colliding-wind binary scenario ($\S$2). The framework is further complicated in D' type systems by the rapid rotation of the cool component which strongly affects the symbiotic environment, leading to a disk-like circumbinary structure where the high gas density enhances grain formation and growth.\\

We have found ($\S$3) that the UV lines are emitted from high density gas between the
stars downstream of the reverse shock, while the optical lines are emitted downstream of
the shock propagating outwards the system. The models which best explain both the observed UV and optical line ratios correspond to \Ts =60,000K; regarding the gas density, in the downstream region of the reverse shock it reaches 10$^8$ \cm3 while it is $\sim$ 10$^6$ \cm3\ downstream of the expanding shock. Free-free radiation downstream of the reverse shock is self absorbed in the radio, so the  data in that range are explained by bremsstrahlung from the nebula downstream of the expanding shock, which becomes optically thick at $\nu \leq$ 5 GHz.

The relative abundances of C, N, and O adopted to reproduce the UV line ratios are in good agreement with those obtained by SN93 for symbiotic stars. Particularly, C/O=0.70 indicates a carbon enrichment of the cool star which can be explained by the transpher of matter from the hot component, a former carbon star before becoming a WD. This hypothesis, suggested by SN93, along with the CIII] 1909/SiIII]1892 ratio $<$ 1 predicted by the models, favour a classification of HD330036 as SS more than as PN. \\

The SED of the continuum ($\S$4) has been disentangled in the different gas and dust contributions: the star fluxes, bremsstrahlung radiation as well as reprocessed radiation by dust from the shocked nebulae. 
Throughout the modelling we have considered silicate grains with \agr=0.2 \mum \, which correspond to the size of grains generally present in the ISM, and large grains with \agr=2.0 \mum \, which are suitable to become crystalline.\\
Three shells are identified in the continuum IR SED, at 850K, 320 K and 200 K  with radii  r = 2.8 10$^{13}$ cm, 4 10$^{14}$ cm, and  10$^{15}$ cm, respectively, adopting a distance to Earth d=2.3 kpc. Interestingly, all these shells appear to be circumbinary.\\
The consistent modelling of line and continuum emission in the shocked nebulae leads to relatively low dust-to-gas ratios, particularly for large grains. Comparing with D-type SS which are generally rich in dust, HD330036 shows $d/g$ lower by factors $>$ 10. Dust reprocessed radiation at $\leq$ 100 K downstream of the shock propagating outwards the system cannot be seen in the SED because hidden by the dust shell radiation flux.\\

The analysis of the ISO-SWS spectrum ($\S$5) has revealed that both PAHs and crystalline silicates coexist in HD330036.
We suggest that the PAHs are associated with the internal shell at 850 K, while crystalline silicates, which derive from annealing of amorphous silicates at temperatures $\geq$ 1000 K, are now stored into the cool shells at 320 K and 200 K.
Strong evidence that crystalline silicates are in a disk-like structure is derived on the basis of the relative band strengths.\\ The proposed scenario would link HD330036 to some bipolar Post-AGB stars which have shown such a dichotomy in
the dust composition, location and geometrical distribution (Molster et al. 2001, Matsuura et al. 2004).\\
The presence of such strong crystalline features is intriguing in the light of our colliding-wind model: as a matter of fact, the temperature required for an effective annealing sufficiently long in order to activate the crystallization process could be reached in the expanding nebula of SS, where colliding winds enable shock waves to compress and heat the dust grains. Indeed the shocks can represent a very suitable mechanism to trigger the crystallization processes, principally thanks to the sudden heating and gradually cooling of the grains in the post-shock region, that might favour the annealing processes. We thus suggest that crystallization processes in HD330036 may be triggered by shocks and that annealing may take place within the circumbinary disk.\\

Our scenario is schematic, of course, and to date should be considered as a mere approximation of the actual physical picture. As a matter of fact, new observations have been revealing the high complexity of these symbiotic environment, where the dynamic of binary components as key parameter is not so trivial to be disregarded entirely.\\
The VLTI/MIDI proposal (P.I. D'Onofrio - 079.D-0242) based on our model and recently accepted by ESO aims to unveil the HD330036 dust environment by means of IR interferometric observations, constraining the morphology and the emitting properties of the PAH-dust shell.
\begin{acknowledgements}
The authors would like to thank A. Cicakova for reading the manuscript.
RA acknowledges the kind hospitality of the School of Physics \& Astronomy of the Tel Aviv University.
\end{acknowledgements}

\end{document}